\documentclass{PoS}

\usepackage{graphicx}
\usepackage{slashed}
\usepackage{color}
\usepackage{amsmath, amssymb}
\usepackage{amssymb}

\title{SU(3) gauge theory with 12 flavours in a twisted box}

\ShortTitle{SU(3) gauge theory with 12 flavours in a twisted box}

\author{\speaker{C.-J.~David~Lin}\\
        Institute of Physics, National Chiao-Tung University,
        Hsinchu 300, Taiwan\\
        E-mail: \email{dlin@mail.nctu.edu.tw}}

\author{Kenji~Ogawa\\
        Institute of Physics, National Chiao-Tung University, Hsinchu
        300, Taiwan\thanks{Address valid to November 2013.}\\
        E-mail: \email{ogawaknj@gmail.com}}

\author{Hiroshi~Ohki\\
        Kobayashi-Maskawa Institute, Nagoya University, Nagoya 464-8602, Japan\\
        E-mail: \email{ohki@kmi.nagoya-u.ac.jp}}

\author{Alberto~Ramos\\
        NIC, DESY Zeuthen, 15738 Zeuthen, Germany\\
        E-mail: \email{alberto.ramos@desy.de}}

\author{Eigo~Shintani\\
        PRISMA Cluster of Excellence, Institut fur Kernphysik and
        Helmholtz Institute Mainz,\\Johannes Gutenberg-Universitt
        Mainz, D-55099 Mainz, Germany\\
        E-mail: \email{shintani@kph.uni-mainz.de}}

\abstract{We present preliminary result for the step-scaling study of
  the coupling constant with the Yang-Mills gradient flow,  in the twelve-favour SU(3)
  gauge theory.  In this work, the lattice simulation is
  performed using unimproved staggered fermions and the Wilson
  plaquette gauge action, from which the gradient flow is also
  implemented. 
  Imposing twisted boundary condition a'la
  t'Hooft and Parisi, our calculation is performed at zero fermion mass.
The renormalised coupling constant is
  extracted via the computation of the energy density.  In order to
examine the reliability of the continuum extrapolation, we investigate
this coupling constant using two different lattice discretisations.
Our result shows that in order to control the systematic effects in
the continuum extrapolation, it is necessary to implement a large enough
gradient-flow time.  In the current calculation, the gauge-field
averaging radius corresponding to the flow time has to
be as large as $40\%$ of the lattice size.  
}

\FullConference{The 32nd International Symposium on Lattice Field Theory,\\
		23-28 June, 2014\\
		Columbia University New York, NY}

\begin{document}

\section{Introduction}
\label{sec:introduction}
Over the past few years, there has been a significant amount of
interest in identifying the minimal number of fermions, $N_{f}^{{\mathrm{cr}}}$,
that results in infrared (IR) conformal behaviour of a particular
gauge theory.   The identification of $N_{f}^{{\mathrm{cr}}}$, which depends on
the gauge group and the fermion representation, can lead to the
construction of candidate models for the composite-Higgs scenario of
electroweak (EW) symmetry breaking.   These models can be realised
using a gauge theory with the number of fermions, $N_{f}$, just below $N_{f}^{{\mathrm{cr}}}$. 

Various approaches have been employed to determine
$N_{f}^{{\mathrm{cr}}}$ for SU(2) and SU(3) gauge theories.   
In all these approaches, one has to work in the regime $N_{f} \sim
N_{f}^{{\mathrm{cr}}}$. Therefore it is challenging to distinguish
between the scenarios of IR conformality and chiral symmetry
breaking.  One of the lasting controversies in this research avenue is
the IR property of SU(3) gauge theory with 12 massless flavours in the
fundamental representation.   While several studies led to the result that
this theory can contain an infrared fixed point
(IRFP)~\cite{Appelquist:2009ty,Hasenfratz:2010fi,Lin:2012iw,Aoki:2012eq,Itou:2012qn,Ishikawa:2013tua,Cheng:2014jba}, there is evidence
to support the opposite conclusion~\cite{Fodor:2011tu}.

In this work, we adopt the method of step scaling~\cite{Luscher:1991wu} to investigate the
behaviour of the running coupling in SU(3) gauge theory with 12
massless flavours.  This method allows us to use the finite lattice size,
$L$, as the renormalisation scale.  The renormalisation scheme, as
detailed in Sec.~\ref{sec:simulation_details}, is defined through
the computation of the energy density with the implementation of
the gradient flow~\cite{Luscher:2010iy,Fodor:2012td}.   The
step-scaling strategy for investigating the IR behaviour of this gauge
theory requires high-precision data.  This is because of the very slow 
running behaviour between the ultraviolet (UV) and the IR regimes.
For SU(3) gauge theory with 12 massless flavours, two-loop
perturbation theory predicts the existence of an IRFP at the
renormalised coupling, $\bar{g}^{2} \sim 10$ ($\alpha_{s} \sim 0.8$),  and
\begin{equation}
  \left [ \frac{\bar{g}^{2}(2L)}{\bar{g}^{2} (L)} \right ]_{{\mathrm{two-loop}}} < 1.07 ,
\end{equation}
everywhere between the asymptotic-freedom point and the IRFP.  Therefore, in
numerical lattice studies, it is
preferable to have data with error in the sub-percentage level, in
order to obtain clear evidence for the existence/non-existence of an
IRFP in this theory.
Furthermore,  it is also necessary to control systematic effects,
which are normally dominated by the errors in the continuum extrapolation,
to the same level.  
This task is very challenging, and was only
partially achieved in previous step-scaling calculations for the
running coupling in this
theory~\cite{Appelquist:2009ty,Lin:2012iw, Ogawa:latt2013}\footnote{In
  Ref.\cite{Cheng:2014jba}, a method similar to the step-scaling
  determination of the coupling constant is used to study SU(3) gauge
  theory with 12 massless flavours.}.
In the present study, the statistical error of our data is well below
$1\%$.  We use two discretisations for computing the same renormalised
coupling.  This allows us to investigate the systematic error resulted from the
continuum extrapolation.   We find that it is necessary to implement a
large enough gradient flow time (with the corresponding gauge-field
averaging radius $\sim 40\%$ of the lattice size) 
in order to control this error.

\section{Strategy and Simulation details}
\label{sec:simulation_details}
We adopt the Wilson plaquette gauge action and
unimproved staggered fermions, with colour-twisted boundary
condition~\cite{'tHooft:1979uj,Parisi:1984cy}.   
Details of the implementation of this boundary condition
in the present calculation are exactly the same as those in
Ref.~\cite{Lin:2012iw}.   
In this work we use the renormalised coupling defined via the
gradient flow. In particular we use the Wilson flow, defined by the
equation (see Ref.~\cite{Luscher:2010iy} for any unexplained notation),
\begin{equation}
 \frac{\partial V_{\mu} (x,t)}{\partial t} = - g_{0}^{2} \left \{
   \partial_{x,\mu} S_{{\mathrm{W}}} \left [ V_{\mu} \right
   ]\right \} V_{\mu} (x,t) , \mbox{ }\mbox{ } V_{\mu} (x,0) = U_{\mu}
 (x) .
\label{eq:WF_equation}
\end{equation}
%

For our step-scaling studies, we use the twisted gradient flow
coupling as defined in Sec. 4 of Ref.~\cite{Ramos:2014kla},
\begin{equation}
  \bar{g}^{2} (L) = \hat{\mathcal N}^{-1}(c_{\tau},a/L) \mbox{ }t^{2} \langle E(t)
  \rangle |_{t=c_{\tau}^{2}
    L^{2}/8} ,
\label{eq:WF_coupling_def}
\end{equation}
where 
\begin{equation}
E(t) = -\frac{1}{2} {\rm tr}  \left ( G_{\mu\nu} G_{\mu\nu} \right )
\end{equation}
is the energy density at
positive flow time, $G_{\mu\nu}$ is the field strength tensor, and 
$c_{\tau}$ is a dimensionless parameter that characterises our
scheme.  Finally $\hat{\mathcal N}(c_{\tau},a/L)$ ensures that at leading 
order our coupling is the same as the coupling in the
$\overline{{\mathrm{MS}}}$ scheme.  We note that 
$\hat{{\mathcal{N}}}(c_{\tau},a/L)$ is computed on the lattice, and 
therefore our coupling definition does not contain any leading order lattice
artefacts.  It is alo free from the zero-mode contribution because of
the use of twisted boundary condition.

Similar definitions of the renormalised coupling have been investigated
in
Refs.~\cite{Cheng:2014jba,Fodor:2012td,Ramos:2014kla,Fritzsch:2013je,Fodor:2014cpa}. 
In these previous works, it was observed that the statistical error of
$\bar{g}^{2}$ are significantly smaller than that in other methods, such as the twisted Polyakov
loop scheme~\cite{deDivitiis:1993hj}.  This is also what we find in
this work, where we compute the coupling every 50 to 200 Hybrid
Monte-Carlo trajectories.

To implement the step-scaling approach, we perform
simulations on lattices with the choices of volume,
\begin{equation}
 \hat{L} \equiv L/a = 6, 8, 10, 12, 16, 20, 24,
\label{eq:all_volumes}
\end{equation}
at many values of the lattice spacing, $a$.   The Wilson flow, as
described in Eq.~(\ref{eq:WF_equation}),  can be shown to result in
the average of the gauge potential in the 4-dimensional sphere with
mean-square radius $\sqrt{8t}$.  This introduces a scale,
\begin{equation}
 \mu = \frac{1}{\sqrt{8t}},
\end{equation}
in the extraction of the coupling 
in Eq.~(\ref{eq:WF_coupling_def}).   Therefore, to implement the
step-scaling method, we have to ensure that the ratio,
\begin{equation}
 c_{\tau} \equiv \frac{\sqrt{8 t}}{L} = \frac{1}{\mu L} ,
\label{eq:ctau_def}
\end{equation}
is fixed in the procedure.  This leads to the definition of a
coupling, through Eq.~(\ref{eq:WF_coupling_def}), at the
renormalisation scale $L$.    In the rest of this article, we will
denote this coupling as $\bar{g}(L)$.
From the
above discussion, it is obvious that $c_{\tau}$ should be larger than
0 and smaller than 0.5.  Picking a value of $c_{\tau}$ corresponds to
specifying a renormalisation scheme.

Tuning the bare couplings on the lattice sizes $\hat{L} = 6, 8, 10,
12$, such that the renormalised couplings are identical, we can make
certain that these lattices are of the same physical size, $L$.  Using
these tuned bare couplings (lattice spacings), we then compute the
renormalised coupling on the lattices that are twice larger.  That is,
we perform the calculation at $\hat{L} = 12, 16, 20, 24$ at these
lattice spacings.  This allows us to carry out the continuum
extrapolation for the coupling renormalised at $2L$.

\section{Analysis and Results}
\label{sec:analysis_and_results}
It is well known that the main source of the systematic effects in the
step-scaling method is the lattice artefacts.
In this work, we use two lattice discretisations, namely the clover
operator and the plaquette, to compute the
energy density in extracting the 
coupling defined in Eq.~(\ref{eq:WF_coupling_def}).  
We denote the ``lattice version'' of the renormalised
coupling generically as $\bar{g}^{2}_{{\rm latt}}$.   By comparing results from
these two discretisations, we can 
investigate the effect of the lattice artefacts and the
reliability of the continuum extrapolation.  Furthermore, we calculate
$\bar{g}^{2}_{{\rm latt}}$ at many values of $c_{\tau}$ between 0 and
0.5.  This allows us to study the effect of $c_{\tau}$ on the size of
the discretisation error.

\begin{center}
\begin{figure}[t]
\includegraphics[width=7.5cm,height=5.2cm,angle=0]{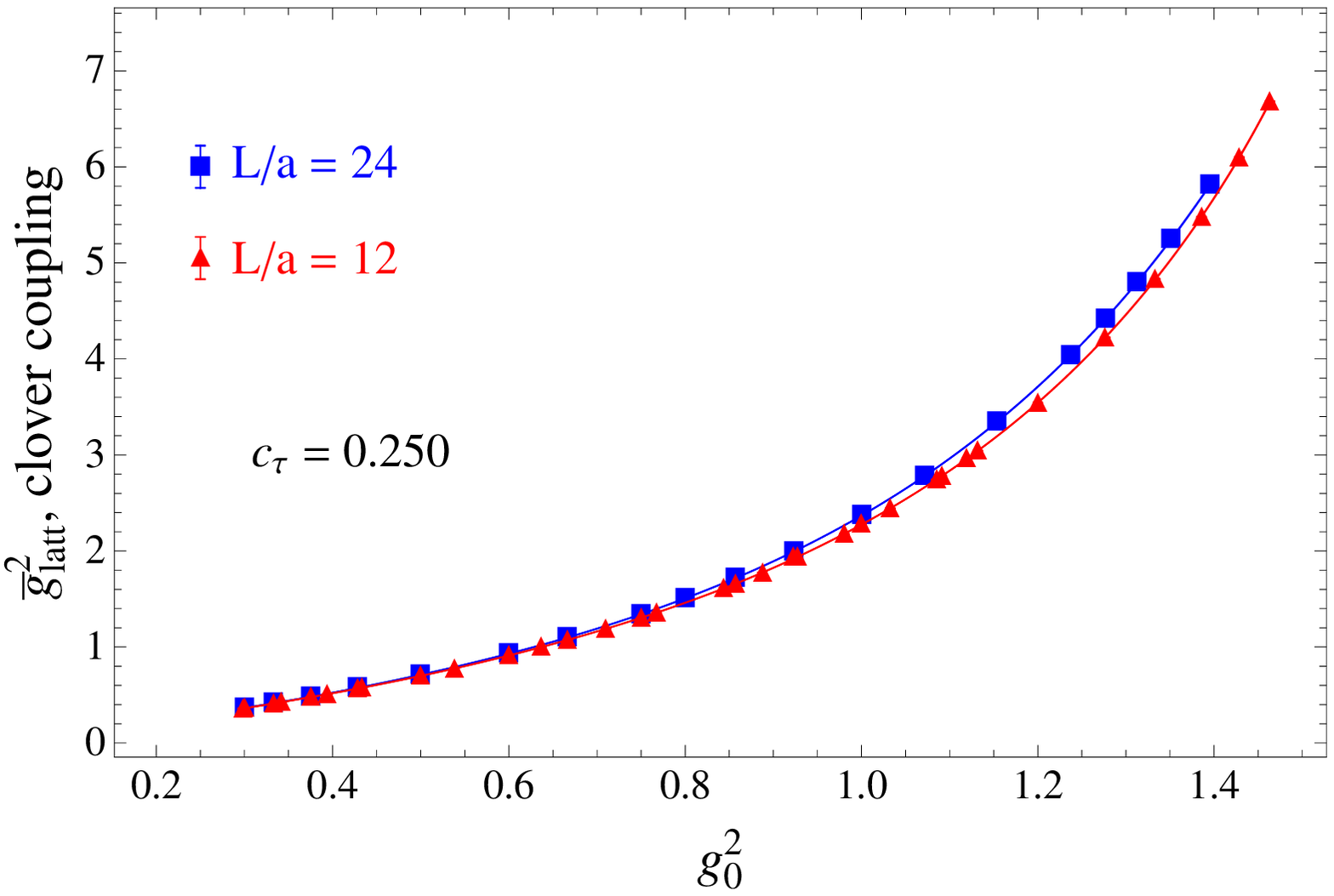}
\hspace{0.5cm}
\includegraphics[width=7.5cm, height=5.2cm,angle=0]{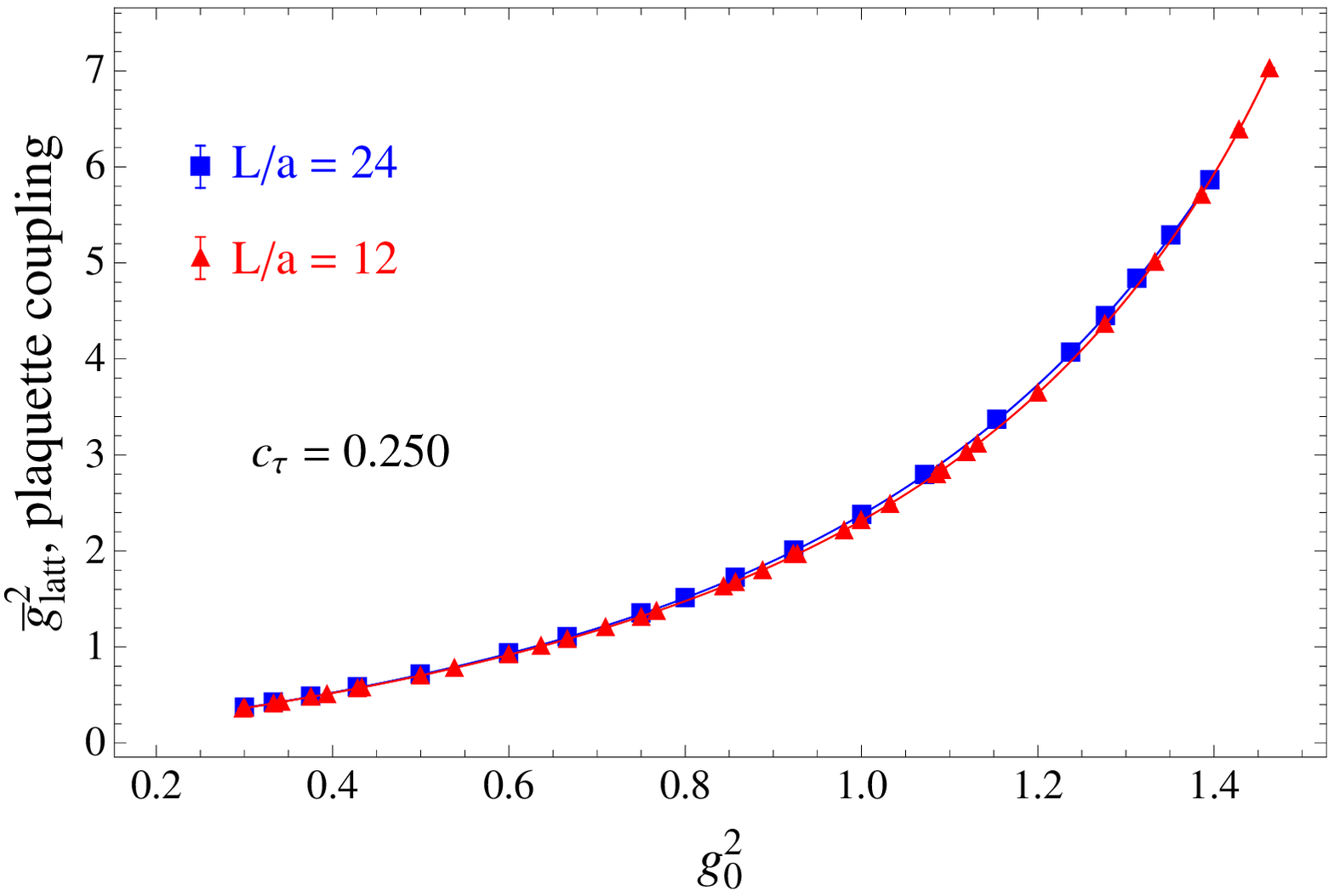}
\includegraphics[width=7.5cm,height=5.2cm,angle=0]{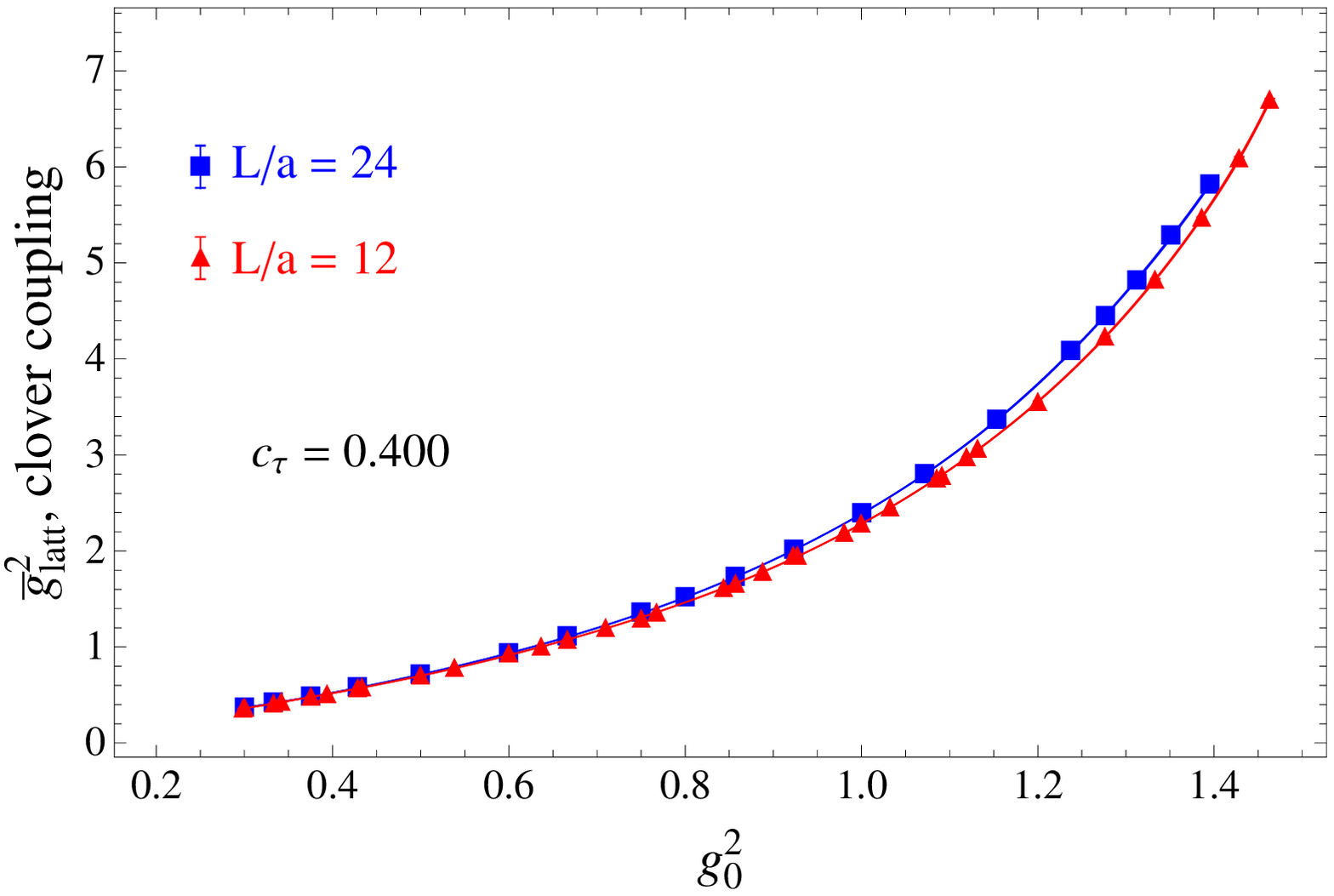}
\hspace{0.5cm}
\includegraphics[width=7.5cm, height=5.2cm,angle=0]{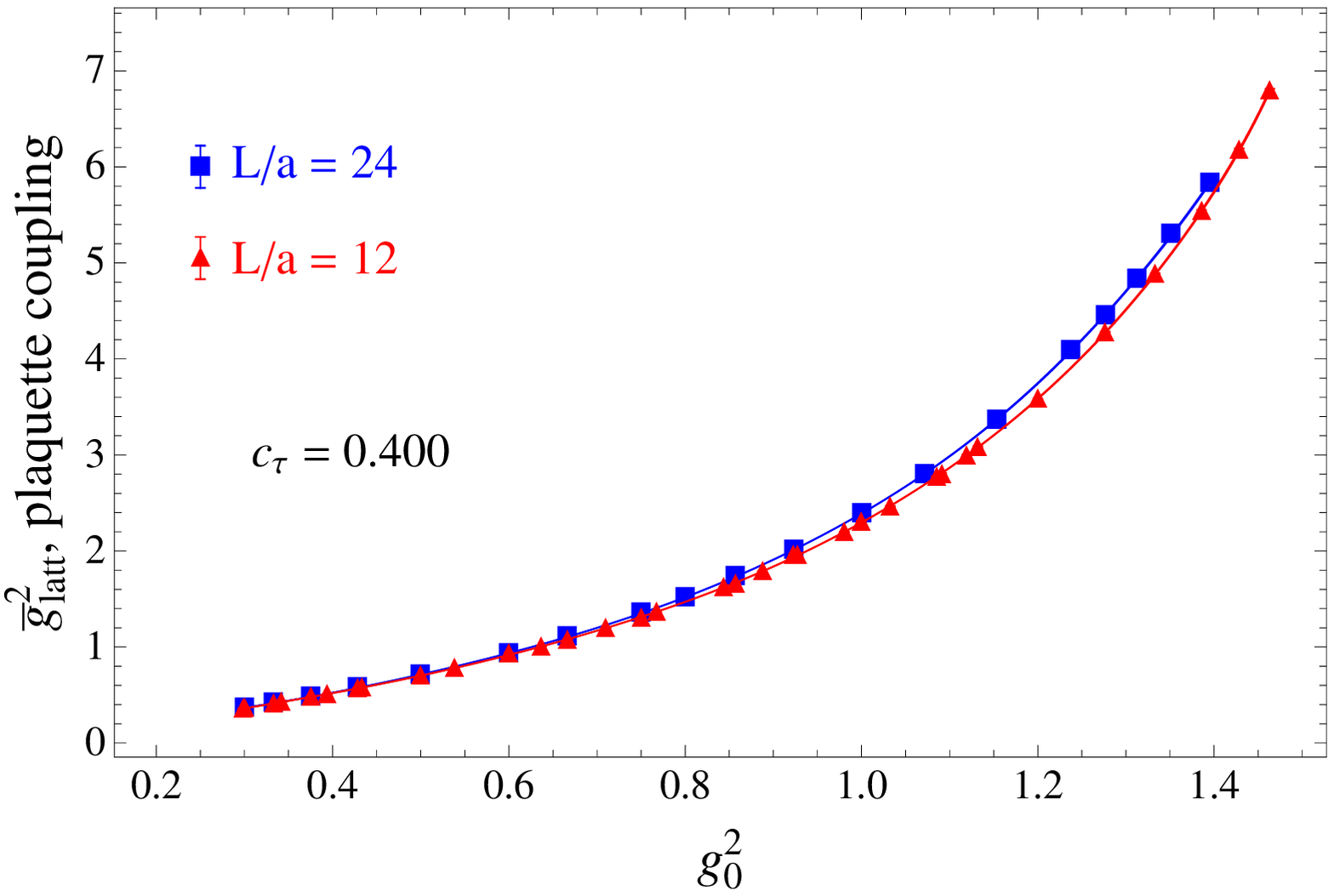}
\caption{\label{fig:beta_interpol}Examples of interpolation in the
  bare coupling in this work.}
\end{figure}
\end{center}
\vspace{-1.1cm}
\hspace{0.68cm}The implementation of the step-scaling approach involves careful
tuning of the bare coupling, in order to set up the constant
physical size of the lattice.  
In practice, we carry out this tuning
procedure by computing the renormalised coupling 
with many values of the bare coupling,
for each lattice volume in Eq.~(\ref{eq:all_volumes}).    At each choice of
$\hat{L}$,  through interpolation, we can then determine
$\bar{g}^{2}_{{\rm latt}}$ as a function of $g^{2}_{0}$ within the interval where we have data.
This interpolation is performed using a non-decreasing
function proposed in Sec. V.B of Ref.~\cite{Lin:2012iw}.
Figure~\ref{fig:beta_interpol} demonstrates the result of such interpolations
for $\hat{L} = 12$ and 24, at two choices of $c_{\tau}$ for both the
clover and the plaquette discretisations.
In the upper right panel, the plot shows the ``crossing'' phenomenon,
which means $\bar{g}^{2}_{{\rm latt}}$ decreases when $\hat{L}$ is
increased at fixed bare coupling constant,
in the strong-coupling regime.  In the current work, this phenomenon
happens for $\bar{g}^{2}_{{\rm latt}}$ computed using the
plaquette discretisation at $c_{\tau} \le 0.45$.  We have not observed it with the clover
discretisation.

\begin{center}
\begin{figure}[t]
\includegraphics[width=7.5cm,height=5.1cm,angle=0]{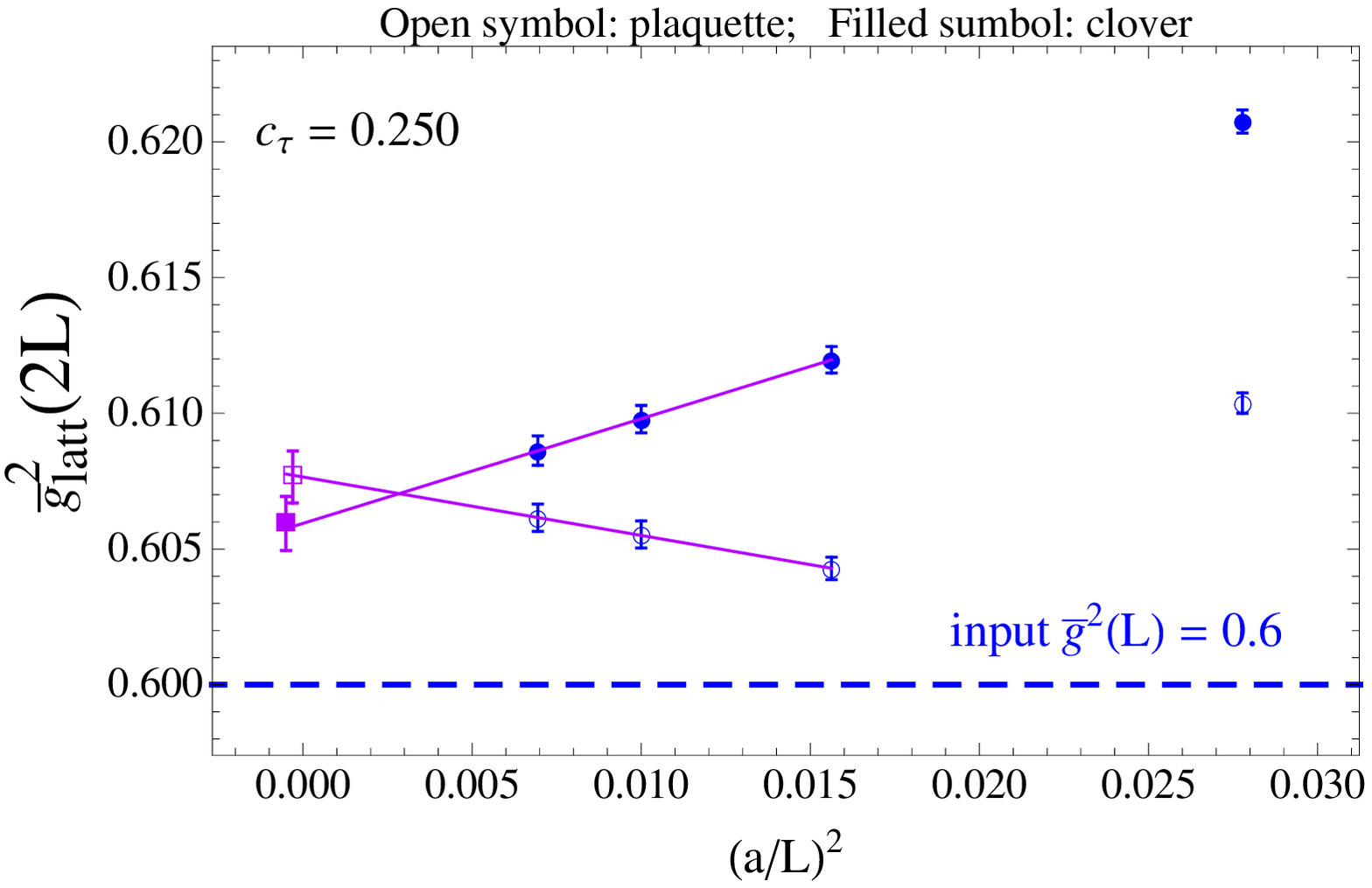}
\hspace{0.5cm}
\includegraphics[width=7.5cm, height=5.1cm,angle=0]{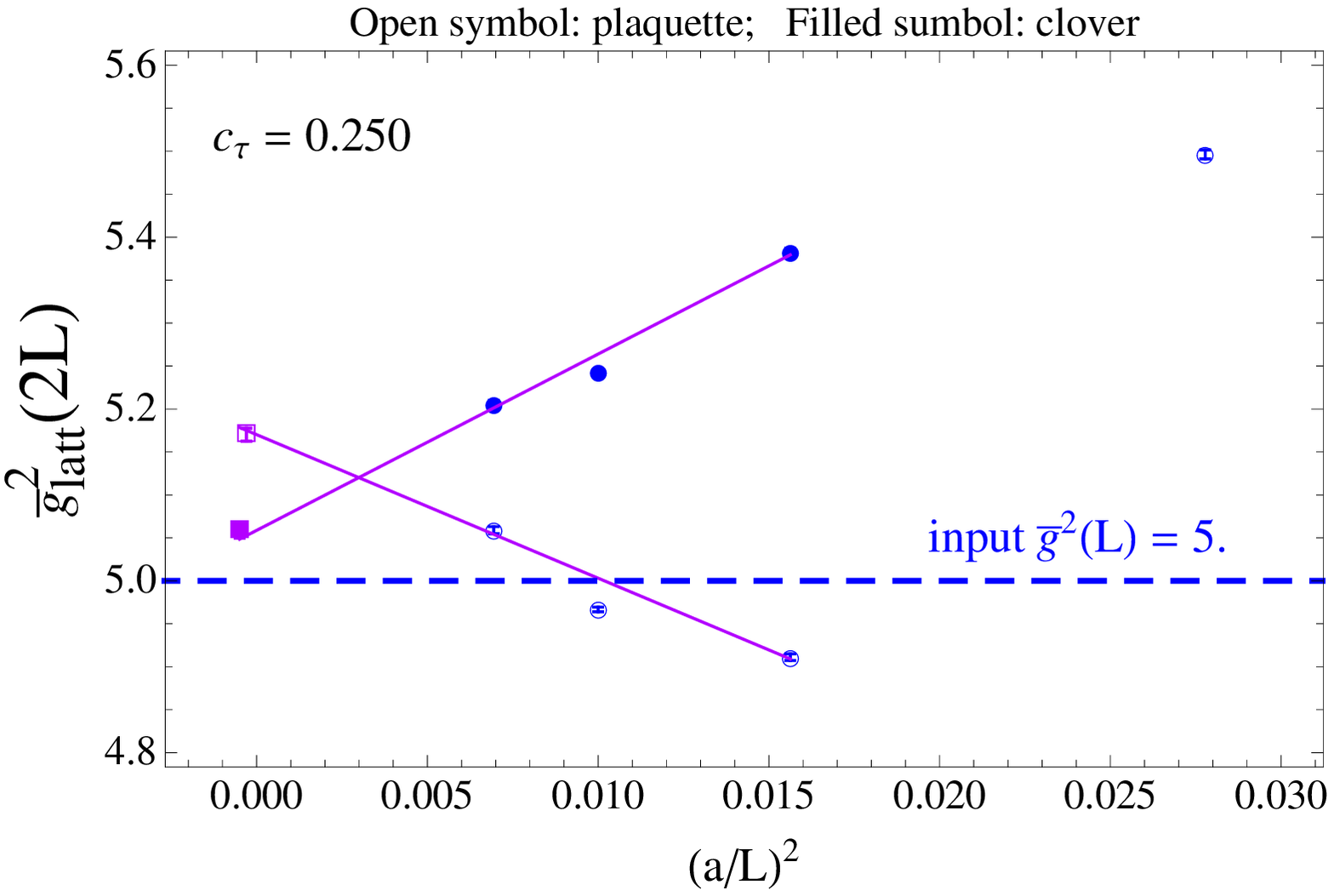}
\includegraphics[width=7.5cm,height=5.1cm,angle=0]{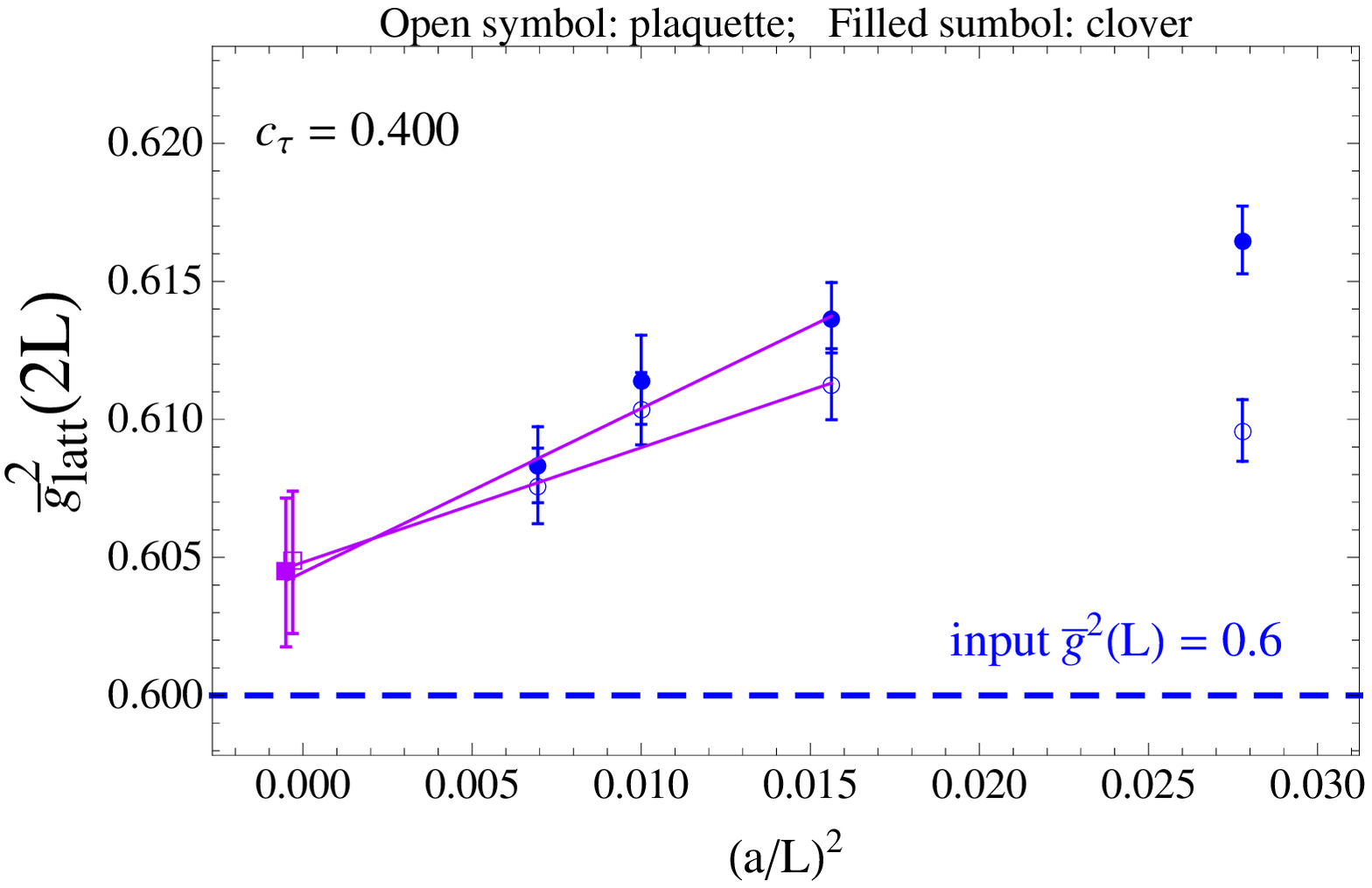}
\hspace{0.5cm}
\includegraphics[width=7.5cm, height=5.1cm,angle=0]{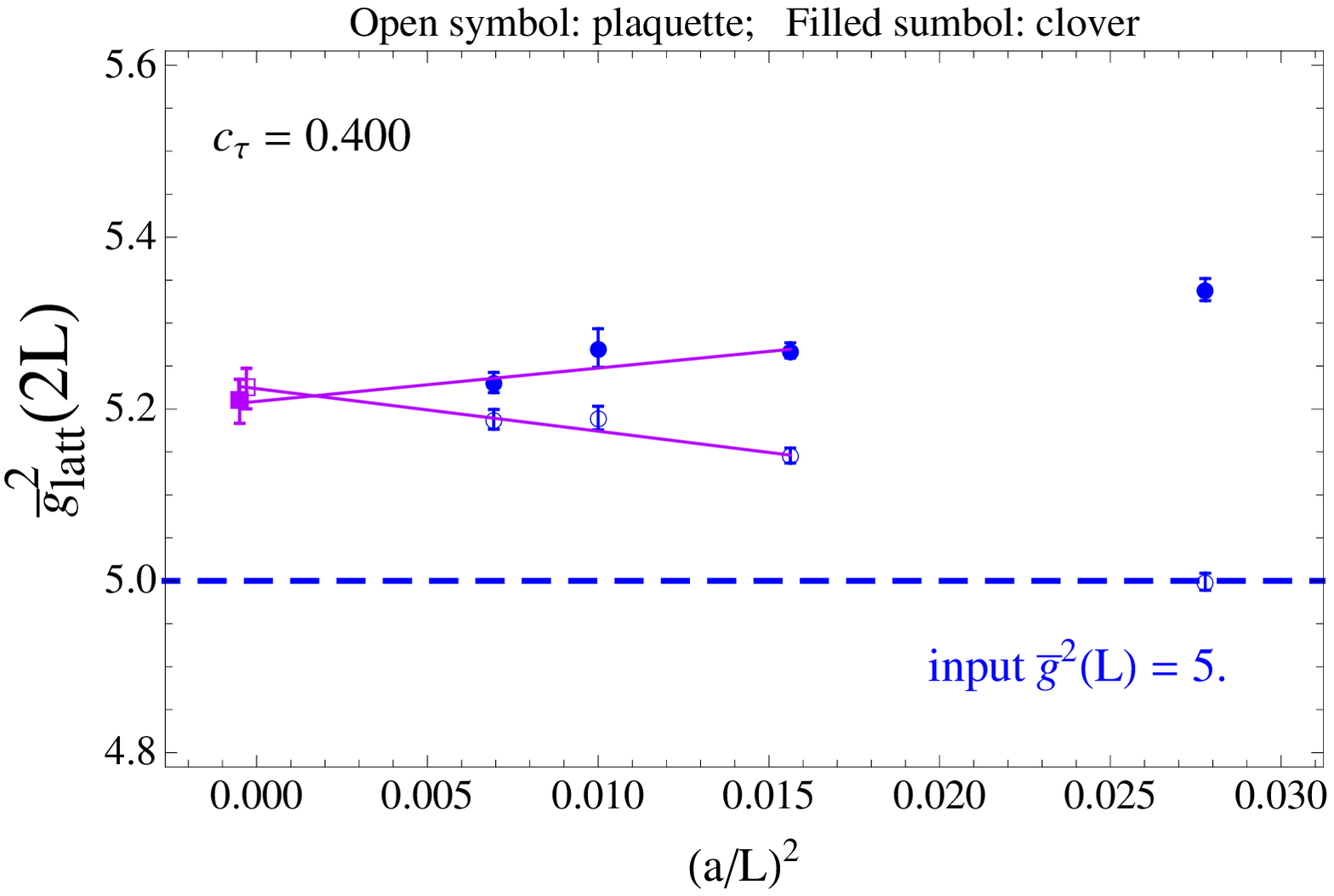}
\caption{\label{fig:cont_extrapol_ctau}Examples of continuum
  extrapolation, using a linear function in $(a/L)^{2}$ and data from the 3
  finest lattices in each case.   In
  the upper-right panel, the result for $\bar{g}^{2}_{{\rm latt}} (2L)$ from the clover discretisation on
  the coarsest lattice is $\sim 6.5$, and is off the range of this
  plot.  The extrapolated results for the plaquette discretisation are
  shifted for clarity.}
\end{figure}
\end{center}
\vspace{-1cm}
\hspace{0.68cm}Once we have performed the bare-coupling interpolation for
$\bar{g}^{2}_{{\rm latt}}$,  it is straightforward to determine the
values of $g^{2}_{0}$
which lead to
\begin{equation}
 \bar{g}^{2} (L) = \bar{g}^{2}_{{\rm latt}} (6 \times a_{6}) =
 \bar{g}^{2}_{{\rm latt}} (8 \times a_{8}) = \bar{g}^{2}_{{\rm latt}}
 (10 \times a_{10}) = \bar{g}^{2}_{{\rm latt}} (12 \times a_{12}) ,
\end{equation}
where $a_{6,8,10,12}$ are the lattice spacings for the lattices with sizes,
$\hat{L}=6, 8, 10, 12$.   This can be done at any $\bar{g}^{2}$ in the
range where we have data.  The next step is to adopt these tuned
$a_{6,8,10,12}$, and perform the continuum extrapolation using
$\bar{g}^{2}_{{\rm latt}} (12 \times a_{6})$, 
$\bar{g}^{2}_{{\rm latt}} (16 \times a_{8})$, 
$\bar{g}^{2}_{{\rm latt}} (20 \times a_{10})$, and
$\bar{g}^{2}_{{\rm latt}} (24 \times a_{12})$.  This enables us to
determine $\bar{g}^{2}(2L)$ for a given value of $\bar{g}^{2}(L)$.  It
is well known that this extrapolation procedure is the main source of
the systematic error in the step-scaling study of the coupling
constant.  This is also the case in our current work.   We first
notice that our coarsest lattice contains significant
lattice artefacts effect.  Therefore we do not use data with
$( \hat{L} = 6 \rightarrow \hat{L} = 12 )$ in the continuum
extrapolation.  This means we carry out a
linear fit in $(a/L)^{2}$ with the data,
\begin{equation}
 \left ( \hat{L} = 8 \rightarrow \hat{L} = 16, \mbox{ }\hat{L} = 10 \rightarrow
   \hat{L} = 20, \mbox{ }\hat{L} = 12 \rightarrow \hat{L} = 24 \right
 ) ,
\end{equation}
for this procedure.
In Fig.~\ref{fig:cont_extrapol_ctau}, we show examples of the 
continuum extrapolation with two choices of $c_{\tau}$.  As expected,
increasing $c_{\tau}$ leads to smaller effects of the lattice
artefacts.   The cases with $c_{\tau} = 0.25$ in the figure show that the
plaquette and the clover discretisations of the same Yang-Mills field
tensor give different results in the
continuum limit, although the individual extrapolation seems
acceptable.  This feature becomes conspicuous in the strong coupling
regime (the upper-right panel of Fig.~\ref{fig:cont_extrapol_ctau}).
Therefore, we conclude that the continuum extrapolation is not under
control at $c_{\tau} = 0.25$.    In fact, we find that it is necessary to
increase $c_{\tau}$ to be $\sim 0.4$, in order to have reliable results
in the continuum limit.  This is demonstrated in the two lower panels
in Fig.~\ref{fig:cont_extrapol_ctau}.

\begin{center}
\begin{figure}[t]
\includegraphics[width=7.5cm,height=5.2cm,angle=0]{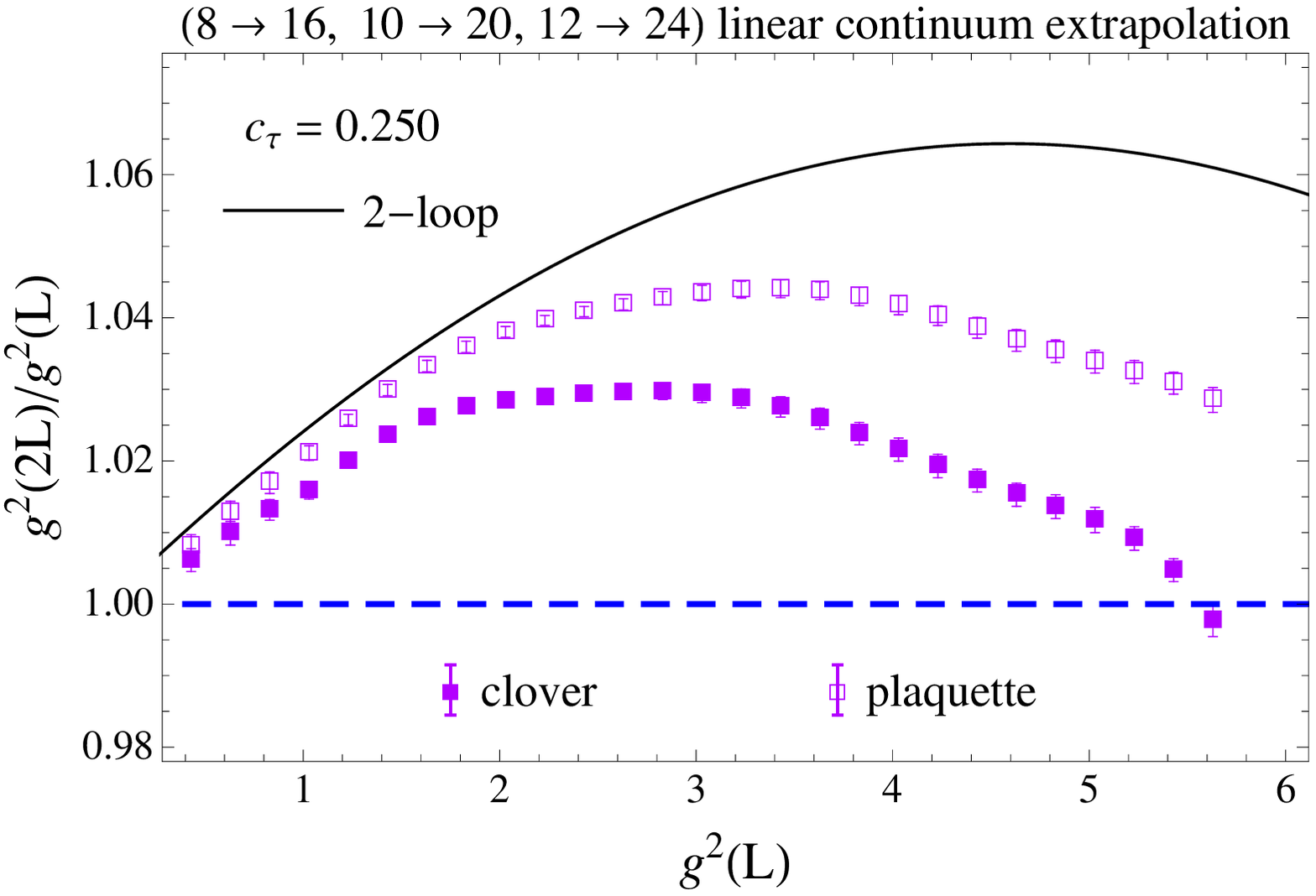}
\hspace{0.5cm}
\includegraphics[width=7.5cm, height=5.2cm,angle=0]{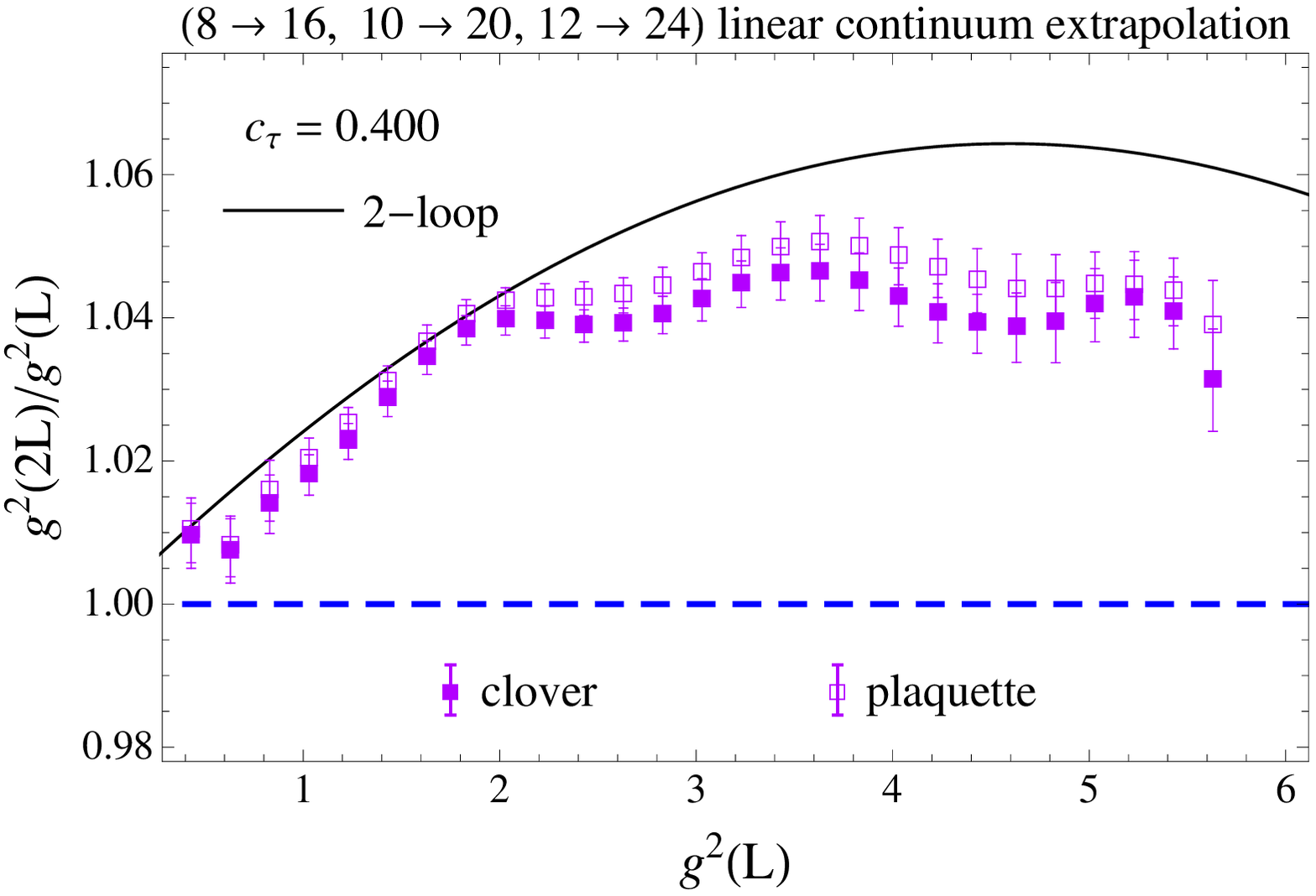}
\caption{\label{fig:step_scaling_result}Results of the step-scaling.
  These plots indicate that the clover and plaquette discretisations
  give consistent continuum-limit result only at large enough
  $c_{\tau}$.  Working with a small value of $c_{\tau}$ and only one
  discretisation method may lead to incorrect conclusion regarding the
  existence of the IRFP.}
\end{figure}
\end{center}

Figure~\ref{fig:step_scaling_result} displays our results for
$\bar{g}^{2}(2L) / \bar{g}^{2}(L)$, plotted against the input
$\bar{g}^{2}(L)$.  
The left panel contains results obtained using the clover and the plaquette discretisations at $c_{\tau} =
0.25$.   This plot shows the importance of investigating the
systematic effect in the continuum extrapolation.   It demonstrates 
that the systematic error in the strong coupling regime is large for 
the case $c_{\tau}=0.25$, such that we cannot draw any conclusion 
regarding the existence of the IRFP.

\vspace{-1.1cm}
The right panel of Fig.~\ref{fig:step_scaling_result} indicates that
in the range of the coupling that we have studied, it is enough to
work with the choice $c_{\tau} = 0.4$ to control the continuum
extrapolation.  Of course, this still has to be confirmed with further
investigation.  Presently we are generating data at one lattice
spacing with $\hat{L}=32$ in the strong-coupling regime, and use it
to examine the reliability of our results in the continuum limit.

\section{Conclusion and Outlook}
\label{sec:conclusion_and_outlook}
In this article, we present preliminary results of our step-scaling
study of the running coupling in SU(3) gauge theory with 12 massless
flavours.  We use the renormalised coupling extracted from the energy
density at positive gradient-flow time with
twisted boundary conditions.   This
renormalisation scheme enables us to have data with statistical errors
at the sub-percentage level.  Such accuracy is essential in the
investigation of the IR behaviour of this theory, in which the
coupling runs very slowly.

We find that cutoff effect in the gradient flow scheme can be
difficult to control. To study this systematic effect, we
use two discretisations to define the coupling. We find
that continuum extrapolations agree for both discretisations only at
large values of the flow time ($c_{\tau} \sim 0.4$).  With various newly-proposed
techniques~\cite{Ramos:latt2014,Sint:latt2014} relevant
to improving the Wilson flow, one may be able to work with a smaller
$c_{\tau}$.  We will study some of these methods in the
future.

Presently we are performing lattice simulations in the regime of
stronger couplings.  We envisage that it is possible to have data up
to $\bar{g} \sim 6.3$ in our work, with simulations still performed on the
weak-coupling side of the bulk
phase structure in the lattice theory~\cite{Kogut:1987ai,Cheng:2011ic}.   It is desirable that our
work will shed light on the controversy over the existence of the IRFP
in SU(3) gauge theory with 12 massless flavours.

\acknowledgments
We are indebted to Tatsumi Aoyama and Hideo Matsufuru for their kind help in developping
the HMC simulation code.  Discussions with Anna Hasenfratz and
Maurizio Piai have been very helpful.  Constant support from Taiwanese National Centre for
High-performance Computing is acknowledged.  We also thank the
hospitality of National Centre for Theoretical Sciences of Taiwan. 
C.-J.D.L. is supported by Ministry of Science of Technology via grant number
102-2112-M-009 -002 -MY3.  H.O. is supported by the JSPS Grant-in-Aid for Scientific Research (S) No.22224003,
and for Young Scientists (B) No.25800139.


\begin{thebibliography}{99}
\bibitem{Appelquist:2009ty}
  T.~Appelquist, G.~T.~Fleming and E.~T.~Neil,
  Phys.\ Rev.\ D {\bf 79} (2009) 076010.
%
\bibitem{Hasenfratz:2010fi}
  A.~Hasenfratz,
  Phys.\ Rev.\ D {\bf 82} (2010) 014506.
%
\bibitem{Lin:2012iw}
  C.-J.~D.~Lin, K.~Ogawa, H.~Ohki and E.~Shintani,
  JHEP {\bf 1208} (2012) 096.
%
\bibitem{Aoki:2012eq}
  Y.~Aoki {\it et al.},
  Phys.\ Rev.\ D {\bf 86} (2012) 054506.
%
\bibitem{Itou:2012qn}
  E.~Itou,
  PTEP {\bf 2013} (2013) 8,  083B01.
%
\bibitem{Ishikawa:2013tua}
  K.-I.~Ishikawa, Y.~Iwasaki, Y.~Nakayama and T.~Yoshie,
  Phys.\ Rev.\ D {\bf 89} (2014) 114503.
%
\bibitem{Cheng:2014jba}
  A.~Cheng, A.~Hasenfratz, Y.~Liu, G.~Petropoulos and D.~Schaich,
  JHEP {\bf 1405} (2014) 137.
%
\bibitem{Fodor:2011tu}
  Z.~Fodor {\it et al.},
  Phys.\ Lett.\ B {\bf 703} (2011) 348.
%
\bibitem{Luscher:1991wu}
  M.~Luscher, P.~Weisz and U.~Wolff,
  Nucl.\ Phys.\ B {\bf 359} (1991) 221.
%
\bibitem{Luscher:2010iy}
  M.~Luscher,
  JHEP {\bf 1008} (2010) 071.
%
\bibitem{Fodor:2012td}
  Z.~Fodor, K.~Holland, J.~Kuti, D.~Nogradi and C.~H.~Wong,
  JHEP {\bf 1211} (2012) 007.
%
\bibitem{Ogawa:latt2013}
  K.~Ogawa, talk presented at Lattice 2013, July 29 - August 3, 2013, Mainz, Germany.
%
\bibitem{'tHooft:1979uj}
  G.~'t Hooft,
  Nucl.\ Phys.\ B {\bf 153} (1979) 141.
%
\bibitem{Parisi:1984cy}
  G.~Parisi,
  in Cargese Summer Institute, 1983, report number LNF-84-4-P, C83-09-01.
%
\bibitem{Ramos:2014kla}
  A.~Ramos,
  arXiv:1409.1445 [hep-lat].
%
\bibitem{Fritzsch:2013je}
  P.~Fritzsch and A.~Ramos,
  JHEP {\bf 1310} (2013) 008.
%
\bibitem{Fodor:2014cpa}
  Z.~Fodor, K.~Holland, J.~Kuti, S.~Mondal, D.~Nogradi and C.~H.~Wong,
  JHEP {\bf 1409} (2014) 018.
%
\bibitem{deDivitiis:1993hj}
  G.~M.~de Divitiis, R.~Frezzotti, M.~Guagnelli and R.~Petronzio,
  Nucl.\ Phys.\ B {\bf 422} (1994) 382.
%
\bibitem{Ramos:latt2014}
  A.~Ramos, these proceedings.
%
\bibitem{Sint:latt2014}
  S.~Sint, these proceedings.
%
\bibitem{Kogut:1987ai}
  J.~B.~Kogut and D.~K.~Sinclair,
  Nucl.\ Phys.\ B {\bf 295} (1988) 465.
%
\bibitem{Cheng:2011ic}
  A.~Cheng, A.~Hasenfratz and D.~Schaich,
  Phys.\ Rev.\ D {\bf 85} (2012) 094509.
%
\end{thebibliography}
\end{document}